\documentclass{article}

\usepackage{arxiv}

\usepackage[utf8]{inputenc} 
\usepackage[T1]{fontenc}    
\usepackage{hyperref}       
\usepackage{url}            
\usepackage{booktabs}       
\usepackage{amsfonts}       
\usepackage{amsmath}
\usepackage{nicefrac}       
\usepackage{microtype}      
\usepackage{lipsum}
\usepackage{graphicx}
\graphicspath{ {./images/} }

\usepackage{tabularx}
\usepackage{xcolor}
\usepackage{fancyvrb}

\title{A multimodal llm for the non-invasive decoding of spoken text from brain recordings}

\author{
 Youssef Hmamouche \\
International Artificial Intelligence Center of \\
Morocco, Mohammed VI Polytechnic University\\
Rabat, Morocco \\
  \texttt{youssef.hmamouche@um6p.ma} \\
   \And
 Ismail Chihab \\
International Artificial Intelligence Center of \\
Morocco, Mohammed VI Polytechnic University\\
Rabat, Morocco \\
  \texttt{ismail.chihab@um6p.ma} \\
  \And
 Lahoucine Kdouri \\
International Artificial Intelligence Center of \\
Morocco, Mohammed VI Polytechnic University\\
Rabat, Morocco \\
  \texttt{lahoucine.kdouri@um6p.ma} \\
 \And
  Amal El Fallah Seghrouchni\\
International Artificial Intelligence Center of \\
Morocco, Mohammed VI Polytechnic University\\
Sorbonne Université, LIP6 - UMR 7606
CNRS, France\\
Rabat, Morocco \\
\texttt{amal.elfallah-seghrouchni@um6p.ma}
}

\begin{document}
\maketitle
\begin{abstract}
Brain-related research topics in artificial intelligence have recently gained popularity, particularly due to the expansion of what  multimodal architectures can do from computer vision to natural language processing. Our main goal in this work is to explore the possibilities and limitations of these architectures in spoken text decoding from non-invasive fMRI recordings. Contrary to vision and textual data, fMRI data represent a complex modality due to the  variety of brain scanners, which implies ($i$) the variety of the recorded signal formats, ($ii$) the low resolution and noise of the raw signals, and ($iii$) the scarcity of pretrained models that can be leveraged as foundation models for generative learning. These points make the problem of the non-invasive decoding of text from fMRI recordings very challenging.  In this paper, we propose and end-to-end multimodal LLM for decoding spoken text from fMRI signals.  The proposed architecture is founded on  ($i$) an encoder derived from a specific transformer incorporating an augmented embedding layer for the encoder and a better-adjusted attention mechanism than that present in the state of the art, and ($ii$) a frozen large language model adapted to align the embedding of the input text and the encoded embedding of brain activity to decode the output text. A benchmark in performed on a corpus consisting of a set of interactions human-human and human-robot interactions where fMRI and conversational signals are recorded synchronously. The obtained results are very promising, as our proposal outperforms the evaluated models,  and is able to generate text capturing more accurate semantics present in the ground truth. The implementation code is provided in \url{https://github.com/Hmamouche/brain_decode}.  
\end{abstract}

\section{Introduction}
At its core, Artificial Intelligence is a discipline aiming to study, break down, and reproduce or mimic some if not all of humanity's cognitive functions. With this scope in mind, it is hard to deny the importance or at least the relevance of Neuroscience to this quest. 

In recent years, integrative research bridging these two disciplines allowed the expansion of this field from classification tasks \cite{sheikh2019decoding,buchweitz2012identifying} into two new types of deeper studies of the encoding and the decoding done in the human brain respectively. The first involves predicting brain activity from conversational signals \cite{yamins2016using,youssef20_interspeech}. This type of research is usually based on external modalities such as speech, gestures, and facial expressions to estimate the underlying neural patterns in the brain. The goal here is to understand multiple forms of how the human brain encodes information, \textit{i.e.}, how specific conversations or stimuli are processed in the brain. In contrast, the second type focuses on decoding conversational signals from brain activity aiming to extract meaningful information from neural signals to reconstruct or interpret the original stimuli without relying on external cues \cite{zhang2022cnn,ozcelik2022reconstruction,takagi2023high,tang2023semantic}. This type of work utilizes advanced machine learning algorithms and neural decoding models to convert neural activity into understandable speech, text, or even reconstructed images and videos.

\begin{figure}[t!]
    \centering
    \includegraphics[width=0.8\textwidth]{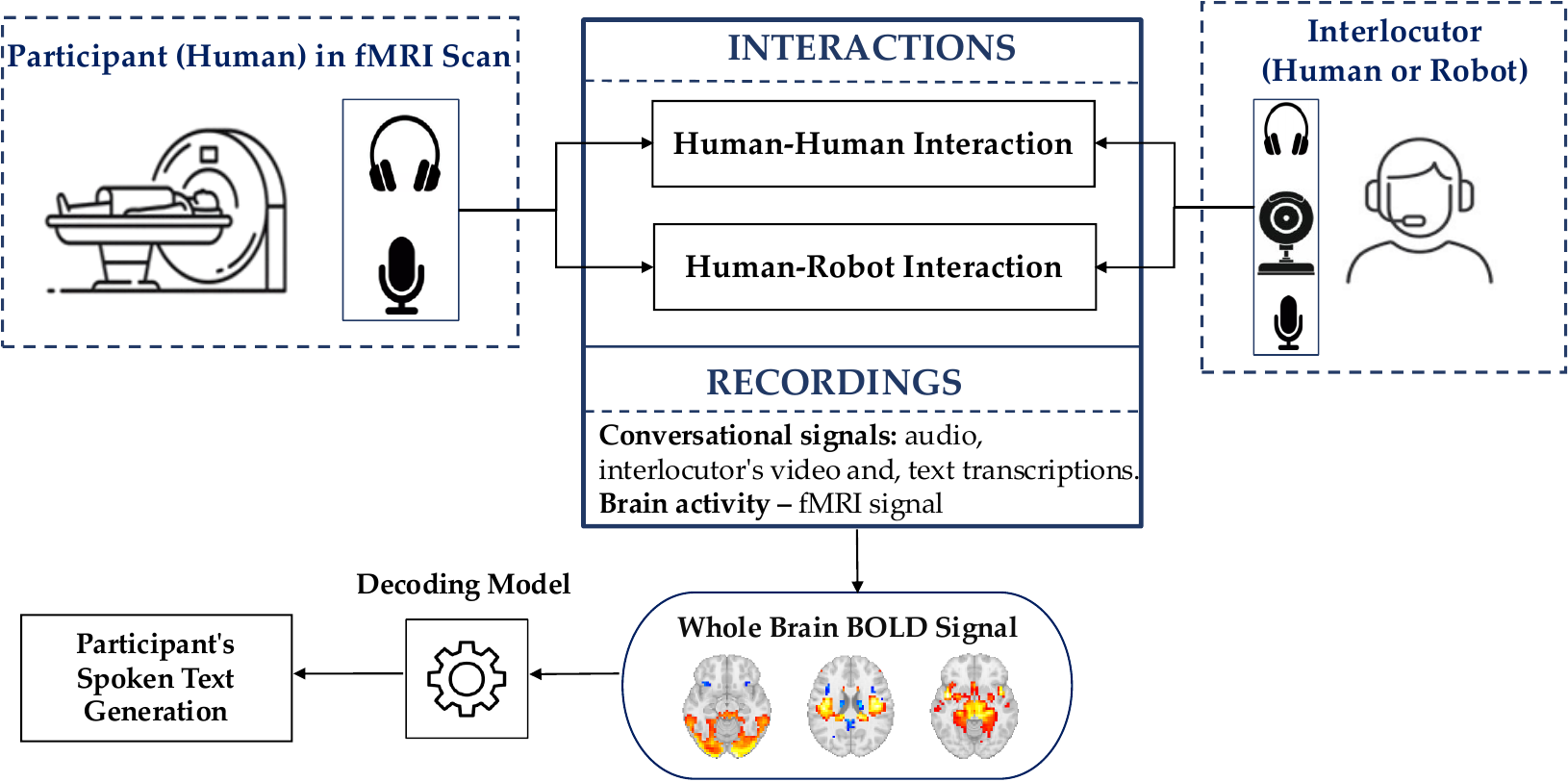}
    \caption{Illustrative diagram of the interaction protocoal in the used dataset \cite{rauchbauer2019brain} and the decoding process of our approach. The datasets consist of conversational and neurophysiological signals recorded during human-human and human-robot interactions. The recorded signals are then leveraged to construct generative models to decode the participant's text from his whole-brain fMRI recordings.}
    \label{fig:interaction_process}
\end{figure}

This decoding process has promising applications in areas such as assistive communication technologies for individuals with speech impairments from brain-computer interfaces enabling direct communication between the brain and external devices, to potentially even mind-reading technologies for improved human-machine interactions. Although its full potential is still being explored, it offers exciting possibilities for bridging the gap between the human brain and external communication.
Research-wise, by estimating and visualizing participants' internal thoughts, such a framework could offer great insight into how the human brain represents information. This would constitute a huge leap forward toward an answer to the fundamental and focal question of neuroscience: How does the human brain work? 
Medically, such an insight and framework could have many applications in fields like realistic emotion recognition and translation for mental and cognitive disorders' treatment from autism and PTSD to even rarer conditions such as Locked-In Syndrome treatment.

Existing approaches rely on generative AI models such as stable diffusion models \cite{ozcelik2023brain,takagi2023high,chen2023seeing}, GANs \cite{NIPS2014_5ca3e9b1,9892673,du2019brain}, Bayesian approach models \cite{tang2023semantic,akamatsu2020multi}, and to a lesser extent transformers \cite{zhang2022cnn}. Although most of these methods yield interesting results, the decoding task here referred to as the reconstruction of original stimuli remains mostly unexplored to its fullest. This is due to two main reasons. The first of which is related to the datasets used for training and testing. These datasets happen to be well-regulated and collected within a measured scope designed for the desired reconstruction task. This defers largely from the much noisier and less regulated real-world scenarios for which such a technology would be used. The second reason, mainly in text decoding as it is the focus of this work, is related to the complexity of the task itself as the reconstruction of clearly presented semantics such as image labeling in \cite{wang2021neural,zhang2022cnn} is a lot less complex than the semantics of a natural conversation. 

In this work, we are primarily focusing on text decoding. Given a set of interactions where conversation and fMRI brain activity are recorded simultaneously between an interlocutor (human or robot) and a participant (human). We propose an end-to-end multimoda model that generates the participant's produced text from a sequence of fMRI recordings of the whole brain. The intuition behind our model is to guess or simulate what participants are talking about based on their brain recordings and what they heard during a conversation. Figure \ref{fig:interaction_process} illustrates the conversations protocol, the recorded signals, and the decoding process. Technically, the proposed architecture connects a pre-trained brain activity encoder and a frozen large language model using embedding alignement. The pre-trained encoder is basically the encoder of a specific transformer that we have adapted to our case with the incorporation of an augmented embedding layer and an improved attention mechanism. The training strategy adopted is carried out in two stages:
\begin{itemize}
    \item Training the transformer to map the text and the associated sequence of brain recordings by drawing a parallel between this task and image captioning.
    \item Connecting the trained-encoder and the frozen LLM using embedding alignment. The alignment is based on a simple projection with feed-forward layer. By leveraging the ability of the pre-trained LLM to understand the language of the target text and its capability to support instruction-tuning, we added the passed interlocutor's text to the LLM as instruction. Given the nature of the data (natural conversations), the intuition behind the proposed system is to behave as a simulator for generating the participant's textual response by taking as input their past brain activity and the text received from the interlocutor (what he listened to).
\end{itemize}

Through an extensive comparative experiment conducted between our proposal and existing architectures, which are mainly inspired by translation-like and image-captioning architectures, the obtained results demonstrate superior results achieved by our system across various text similarity and semantic metrics. Qualitative results further illustrate the visual quality of the conversation context decoded by our model, which identifies key conversational keywords in many examples unlike other models. Therefore, this work validates the capabilities of  multimodal large language models and leads the way for their utilization in the task of text decoding from brain recordings during conversations

After presenting related work in the next section, we describe the proposed approach in Section \ref{sec:approach}, the experiments and results in Section \ref{sec:results}, and a discussion in Section \ref{sec:discussion}.

\section{Related Works}
Although our main focus in this study revolves around the decoding of linguistic semantic signals, particularly text, from fMRI brain activity, we also draw inspiration from image decoding. This choice is motivated by the commonality of the fMRI processing phase in the two tasks. Accordingly, we begin our discussion by briefly exploring approaches related to decoding images from brain recordings before moving on to text decoding.

\subsection{Decoding Images from Brain Recordings}
Drawing inspiration from conditioned GANs and style GANs, many methods use the natural images as a constraint on the fMRI embedding space as well as ground truths to train the model \cite{ozcelik2023brain,takagi2023high,ozcelik2022reconstruction}. Ozcelik et al. \cite{ozcelik2022reconstruction} achieves this by splitting the latent space into three variables, training three regression models to each predict one of these variables from the corresponding fMRI, and using the combination of these regression models, the latent variables they predict to condition a BigGAN architecture to reconstruct the original images.

Based on the same intuition, ren et al. \cite{ren2021reconstructing} distinguish between two encoders, one for the fMRI scans and the other for the natural images, $E_{cog}$ and $E_{vis}$ respectively. Then introduces a teacher-student relationship between them through the decoder. $E_{vis}$ is trained alongside the decoder in a variational autoencoder paradigm. The decoder's weights are then frozen to train the $E_{cog}$ the same way. This ensures that during its training, the latent space $E_{cog}$ (\textit{i.e.}, the cognitive encoder) constructs is close to $E_{cog}$ (\textit{i.e.}, the visual encoder) and is hence better adapted for the task. 

\subsection{Decoding Text from Brain Recordings}

Decoding text from human brain activity is a subject that has been widely investigated over the last two decades using different types of recordings (Functional magnetic resonance imaging (fMRI), Electroencephalogram (EEG), or Magnetoencephalography (MEG)).
The primary motivation for this task was to develop a method for understanding how the human brain processes information and generates thoughts and behaviors. This could be used to map the brain to text in order to identify brain areas involved in text processing. The main techniques used were based on classical machine learning models and based on correlation and regression models \cite{pereira2018toward,herff2015brain}.

Recently, and because of the big scale-up in power for generative models, the work around this task has experienced a shift in paradigm too. Most recent approaches either adopt end-to-end networks and generate text and images directly from the activity of the whole brain \cite{zhang2022cnn,luo2022brain} or go around the weakness of traditional machine learning algorithms by adding a pre-trained backbone to the architecture\cite{tang2023semantic}.
Zhang et al. \cite{zhang2022cnn} present an example of the end-to-end approaches, based on a CNN-transformer hybrid architecture designed to decode a descriptive sentence of the visual stimuli from fMRI brain activity. The stimuli consist of linguistic and visually perceived signals. The proposed architecture is based on a self-attention-based transformer leveraging the ability of these networks in text translation. The input of the model is a sequence of brain recording hence the use of the convolution layer at the beginning of the encoder network to extract spatio-temporal features from the input. The second modification compared to the classical transformer is the introduction of multi-layer connectivity through the encoder and decoder layers in the form of a scaled-attention mechanism. The output from the last encoder layer, usually used to calculate the encoder-decoder attention in the decoder, is replaced by a weighted sum of the outputs of all the encoder layers.
On the other hand, tang et al. \cite{tang2023semantic} present a pre-trained large language model with a combination of classical machine learning algorithms that can yield great results in this decoding task. The authors propose a new methodology to decode continuous text (long phrases, word by word) from brain activity. The approach here is different, the network's encoder uses regression to map listened words and the corresponding the Blood-Oxygen-Level-Dependent (BOLD) signal. The decoder is an English-based language model, \textit{i.e.}, GPT1  trained on another large English Dataset. It is responsible for predicting candidates for the next word in each sequence connected with the encoder to select the most likely next word using a beam search algorithm.
The authors also discuss the potential source of decoding error as mainly related to the resolution of  BOLD signal.

\begin{figure}[t]
\centering
\includegraphics[width=1\textwidth]{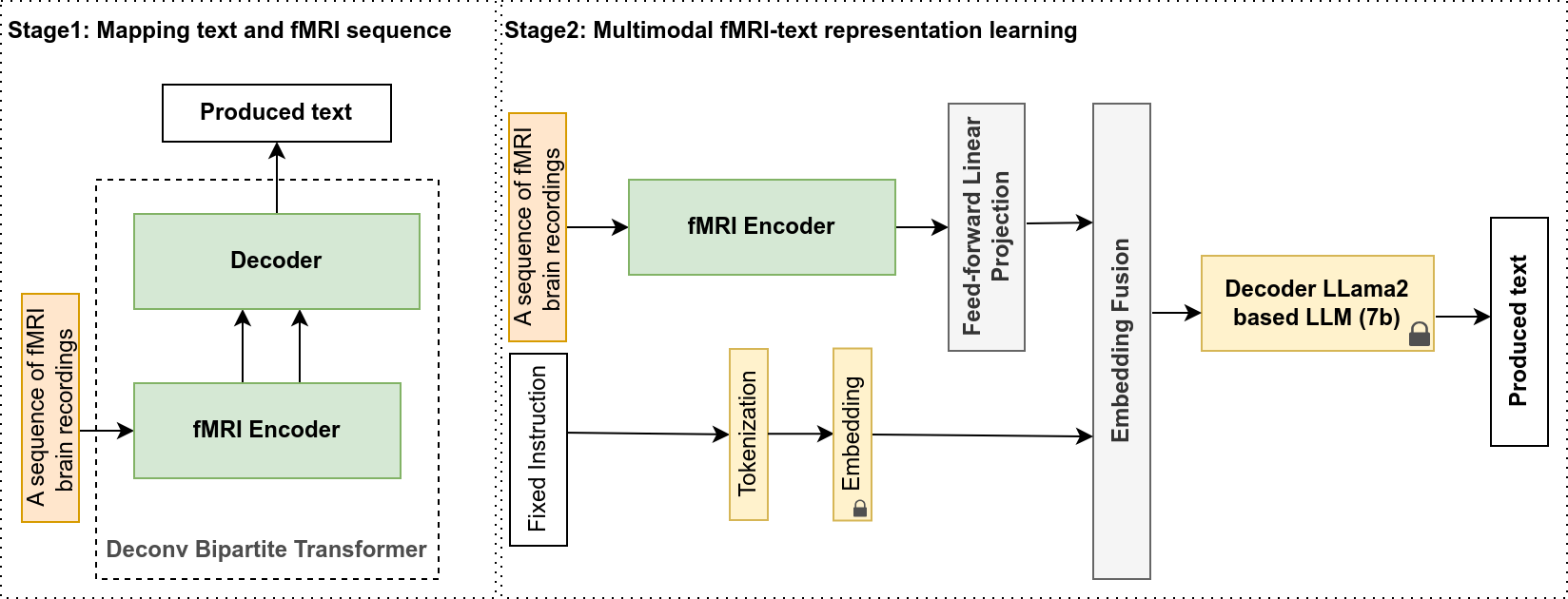}
\caption{Schema of the proposed system -  fMRI brain activity to text representation learning via two training stages: (1) brain activity and text mapping with an improved deconvolution bipartite transformer, and (2) multimodal generative pre-training using the trained encoder and a frozen large language model as a decoder.}
\label{fig:framework}
\end{figure}

\section{Method}
\label{sec:approach}
In this section, we present the proposed method.
Our decoding task can be formulated as follows:
Given a set of interactions where conversation and fMRI brain activity are recorded simultaneously between an interlocutor (human or robot) and a participant (human), the goal is to predict the participant's produced text from a sequence of fMRI recordings of the whole brain. In other words, we want to guess what the participants are talking about from their brain recordings. 

To decode text from a sequence of fMRI recordings, we draw a parallel between this reconstruction task and the well-known task of machine translation. 
In this context, the transformer architecture is widely used for text generation. However, it is generally used for text-to-text generation.
The idea here is to adapt this architecture to work with two different modalities. The encoder takes as input a sequence of fMRI recordings (whole brain) and the decoder outputs the text, \textit{i.e.}, the participant's produced text during the conversation.
A variation of the classic Transformer Architecture has been proposed in \cite{zhang2022cnn}. By modifying the classical text embedding layer in the encoder, and the fully connected layers in both the encoder and decoder with 1D Convolution operations, this work successfully adapts the transformer architecture for this multimodal task.  
The drawback of this approach is that it seeks to adapt the architecture to work with the data without considering the specificity of the fMRI signals, which are usually noisy and have a low frequency compared to electroencephalogram (EEG) and Magnetoencephalography (MEG) for instance.

 In the rest of this section, we detail our contribution that consists of two architectures: (1) An improved transformer to decode text from fMRI recordings by following the methodology of existing approaches, and (2) a new architecture based on a multimodal LLM that employs the proposed transformer encoder as a foundation model.

\subsection{A Multimodal LLM}
The intuition here is to put a multimodal LLM at the place of the participant and training it to imitate his textual response during conversations given the recorded brain activity and the pereceived text from the interlocutor. For this purpose, we combine three concepts: (1) translation-like transfomers for their ability to map representations via encoder-decoder architectures,  (2) LLMs for their capability in generating human-like text, and (3) alignement techniques to allow frozen LLMs supporting and understanding multimodal representations.

The proposed system is illustrated in Figure \ref{fig:framework}. It combines the pretrained encoder of the proposed Tranformer and a frozen Llama-2 (7b) based LLM.  The training strategy is carried out in two stages:
\begin{itemize}
    \item Training the transformer to map the text and the associated sequence of brain recordings.
    \item Connecting the trained encoder and the frozen unimodal LLM using embedding alignment. The alignment is based on a simple projection with a feed-forward layer. 
\end{itemize}

\subsection{A bibartite Transformer with inception deconvolution-based fMRI encoder}
\begin{figure*}[t]
    \centering
    \includegraphics[width=1\textwidth]{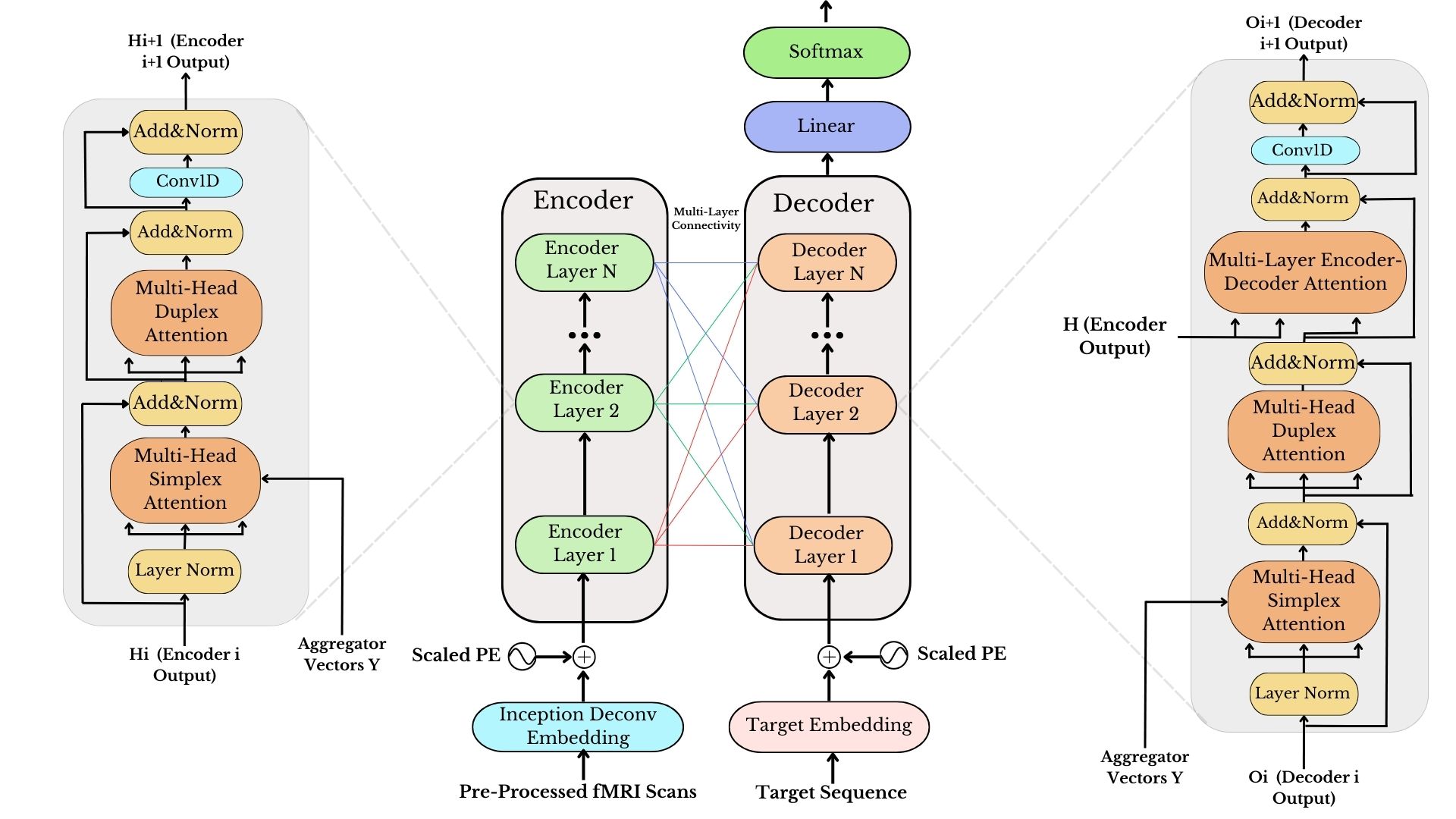}
    \caption{Architecture of the proposed Bipartite Transformer for text decoding from fMRI brain activity.}
    \label{fig:model}
\end{figure*}

Here, we introduce an new method for representing fMRI signals within the Transformer model, specifically focusing on the initial segment of the encoder. Given the low resolution of the BOLD signal, instead of using convolution to extract features from the input signal as used in \cite{zhang2022cnn}, we employ deconvolution using multiple filters in parallel using the inception mechanism to augment the temporal resolution of the input signal and extract more representative features. In addition, we use a specific attention mechanism, as proposed in \cite{hudson2021generative}, in the encoder and decoder of our Transformer to take into account the complexity of the task studied. This attention mechanism introduces a second parameter to the function: a set of latent or aggregator variables. Formally, bipartite attention allows communication between two sets of variables, $X^{n \times d}$ the input data, fMRI signals in our case, and $Y^{n \times d}$ a set of trainable aggregator vectors. This communication can be done in two different ways: {\bf{Simplex Attention}} and {\bf{Duplex Attention}}. The proposed architecture is illustrated in Figure \ref{fig:model}.

In the remainder of this part, we explain the first deconvolution embedding layer and the simplex and duplex attention mechanisms utilized in our proposed architecture. Lastly, we outline the complete bipartite transformer.

\subsubsection{Inception Deconvolution Embedding}
The key concept behind this block is the use of "Transpose Inception Modules" which consist of a set of parallel 1-dimensional transposed convolutional layers with different filter sizes. Specifically, we define three Deconvolution branches each with three kernel size (($1\times1$), ($3\times3$), and ($5\times5$)), an input channels parameter equal to the feature dimension of our fMRI input data, and an output channels parameter equal to the inner dimension of our model, followed by a ReLU activation. These different filter sizes allow the network to capture features at different scales, enabling it to detect both fine-grained and coarse-grained patterns in the input fMRI while the change in channels reduces the data's dimensionality to match the model's input. This helps the model to learn more abstract and higher-level representations of the input data along the temporal dimension. The outputs from these parallel paths are then concatenated, hence expanding the time steps dimension of our fMRI signals into a richer representation.

\subsubsection{Simplex Attention}
This type of attention distributes information in one direction. It lets the trainable aggregator vectors Y alter the statistical distribution of X according to the following rule: 
\begin{equation}
u^s(X, Y)=\gamma(a(X, Y)) \odot \omega(X)+\beta(a(X, Y)),
\end{equation}
\noindent where \noindent$a(., .)$ stands for the classic attention mechanism calculated as:
\begin{equation}
a(X, Y)=\text{Attention}(q(X), k(Y), v(Y)),
\end{equation}

\noindent where $\gamma(.)$ and $\beta(.)$ are trainable mappings that introduce gain and bias to the classic attention score. $\omega(.)$ is a normalizing factor calculated as: \begin{equation}
\omega(X)=\frac{X-\mu(X)}{\sigma(X)}.
\end{equation}

\subsubsection{Duplex Attention}

In this form of attention, the flow of information goes in both ways. The statistical distribution of the input fMRI signals and the aggregator vectors both affect each other. This is done by assuming $Y^{n \times d}$ has a key-value structure and hence accepts its own attention score. 
With this new structure, the update rule becomes:

\begin{equation}
u^d(X, Y)=\gamma(A(Q, K, V)) \odot \omega(X)+\beta(A(Q, K, V)),
\end{equation}
where $\gamma(.)$, $\omega(.)$ and $\beta(.)$ are the same as before.

\subsubsection{A Bipartite Transformer}

Although it is based on the same attention mechanism introduced in \cite{hudson2021generative}, their architecture utilizes this attention in a StyleGAN Architecture to generate images. Our architecture merges this attention mechanism with the general architecture of the CNN-Transformer Hybrid presented in \cite{zhang2022cnn}.

\begin{itemize}
    \item[-] {\bf{Scaled Positional Embedding:}} To account for the temporal modality of fMRI signals and its significant importance in comparison to the positional patterns of an input sentence in a classic transformer architecture, we adopt the same Scaled Positional Embedding as \cite{zhang2022cnn}. The main difference is the introduction of two trainable weights for the positional embedding of both the encoder and the decoder's input, respectively $\lambda$ and $\eta$. 
        \begin{equation}
            \begin{aligned}
            & x_i=\tilde{x}_i+\lambda P E(i), i=1,2, \ldots, n \\
            & y_i=\tilde{y}_i+\eta P E(i), i=1,2, \ldots, m
            \end{aligned}
        \end{equation}
    \item[-] {\bf{Encoder Blocks:}} Different from the classic Transformer architecture, our model's encoder block counts three inner components. The first is a Simplex attention block, followed by a Duplex attention block and a feed-forward block composed of two Conv1D layers separated by a ReLu Activation function.
    \item[-] {\bf{Multi-Layer Encoder Decoder Attention:}} This attention ($Attention_m$) differs from that of the classical transformer Decoder in that it attends to a stack of the outputs from all the encoder blocks as the weighted sum of \(N\) standard attention mechanisms \(Attention(Q, K, V)\), as follows:
    \begin{equation}
    Attention_m = \sum_{i=1}^{N} \alpha_{i} \cdot Attention(W_{q}Z,W_{k}H_{i} ,W_{v}H_{i})
    \end{equation}
    and
    \begin{equation}
    \alpha_{i} = \sigma (W_{i}[Z, Attention(W_{q}Z,W_{k}H_{i},W_{v}H_{i})] + b_{i}),
    \end{equation}
    where $\alpha_i$ denotes the weight of the standard attention mechanism Attention(Q, K, V) computed from the output Z of the previous decoder module and the output Hi of ith encoding layer, $Q = W_qZ$, $K = W_kH_i$ and $V = W_vH_i$, where [,] is a concatenation operation, $\sigma$ is the sigmoid function, $W_q$, $W_k$, $W_v$, $W_i$ and $b_i$ represent the trainable weights.
    \item[-] {\bf{Decoder Blocks:}} In the same way, the decoder block in our architecture is composed of four inner blocks. The first two are respectively Simplex and Duplex attention followed by a multi-layer Encoder- Encoder-Decoder attention and the same feed-forward block as the encoder.
\end{itemize}

\section{Experiments and Results}
In this section, we provide a benchmark involving two established state-of-the-art architectures alongside our proposals.
To justify the effectiveness of our architecture, we add three variants of it as an ablation study, each tested without a particular block to evaluate its impact. 

\label{sec:results}
\subsection{Experimental Setup}
To account for the added attention layer with each Transformer Block in our proposed model, we trained our model with N=2 Transformer Layers.
The softmax cross-entropy is used as the loss function of the proposed decoding models. The Hyperparameters $d_{model}$, $d_{ff}$, $h$, and $d_k$ were respectively set to 256, 512, 8, and 32.  The Adam Optimizer is used with an initial learning rate of $10^{-4}$, and a maximum of 300 training epochs, converging within 200 epochs for Transformers and 300 epochs for MLLMs.
The experiments are conducted in a single accelerated by an NVIDIA GPU A$100$.

\subsection{Dataset Description}
The dataset used in this work is presented in \cite{rauchbauer2020multimodal}. This corpus examines real-life, bidirectional conversations of Human-human and Human-robot Interaction of twenty-five native French speakers in four sessions. The participants are told that the study's aim is to test new key messages for an advertising campaign. In each session, the participants are shown an image of six images that define the general context of the experiment. The subject then engaged in six conversations of $60$ seconds each (three with a human and three with a conversational robot) during which the subjects's fMRI signals and the conversation were recorded. The resulting corpus is composed of three modalities: 
\begin{itemize}
    \item[-] {\bf{The 6 stimuli images:}} Shown to the subjects before the fMRI scan, these images are of an eggplant, a lemon, and an apple carved like Batman, Ninja Turtle, and Spiderman respectively, and of a Rotten Strawberry, Pear, and Rasberry.
    \item[-] {\bf{Pre-processed fMRI scans:}} These scans are in the numerical form of 53 time-steps (1.205 seconds each) and 274 features (atlas clusters) representing the BOLD signal in each of the relevant brain areas.
    \item[-] {\bf{Conversation Transcripts:}} Praat file transcriptions of the recorded conversations.
\end{itemize}
Each of these modalities comes in its own separate file containing exactly 594 samples.

\begin{table}[!ht]
\caption{Example of one-minute conversation from the used dataset in original French (in blue italics) with segment durations in seconds. The context of this conversation is conditioned by an image of a Raspberry shown to the participant before starting the conversation.}
\centering
\begin{tabularx}{0.99\textwidth}{|p{2cm}|X|}
\hline
\textbf{Participant:} (1.31, 3.19) &  \color{black} {so this was a raspberry.} \color{blue} \textit{donc là il s'agissais d'une framboise.}    \\ \hline

 \textbf{Interlocutor:} (3.08, 3.25) &   \color{black} {yeah.}  \color{blue} \textit{ouais}. \\ \hline
 
 \textbf{Participant:} (3.82, 14.73) &   \color{black} {that wasn't, uh, uh, real, in which we'd cut flesh, but rather a drawing, uh, here's a raspberry with eyes.}  \color{blue} \textit{qui faisait non pas euh, euh réelle dans laquelle on avait, coupé de la chair, mais là il s'agissait plutôt d'un dessin, euh, voilà euh une framboise avec des yeux.}    \\ \hline
 
 \textbf{Interlocutor:} (14.60, 15.47) & \color{black} {ah yes okay yes.} \color{blue} \textit{ ah oui d'accord oui.}   \\ \hline
 
 \textbf{Participant:} (14.93, 16.11) &  \color{black} {feet, hands.} \color{blue} \textit{des pieds des mains.}    \\ \hline
 
 \textbf{Interlocutor:} (16.53, 24.28) &   \color{black} {but uh it was more computer- uh well it wasn't in continuity there we are more in the superheroes with the beautiful fruits and vegetables with the cut-out things eh there it's uh.} \color{blue} \textit{mais euh c'était plus ordi- euh enfin c'était pas dans la continuité là on est plus dans les super héros avec les beaux fruits et légumes avec les trucs découpés hein là c'est euh. } \\ \hline
 \textbf{Participant:} (22.40, 25.75) &  \color{black} {that's it, plus it would have said.}  \color{blue} \textit{ c'est ça, en plus on aurait dit.}  \\ \hline

 \textbf{Interlocutor:} (25.04, 29.46) & \color{black} {so yeah it's a drawing you say so it's not a real one it's not something that would have been uh.} \color{blue} \textit{donc ouais c'est un dessin tu dis donc c'est pas une vraie c'est pas quelque chose qui aurait été euh.}   \\ \hline

 \textbf{Participant:} (29.88, 41.50) &  \color{black} {no, that's it, we would have said something, rather a computer, but uh well she had a, funny expression, this raspberry as if uh, she had been a little, I don't know, she seemed upset, this raspberry.} \color{blue} \textit{non c'est ça on aurait dit quelque chose voilà plutôt un ordinateur, mais euh bon elle avait une, drôle d'expression cette framboise comme si euh, elle avait été un peu, je sais pas elle semblait contrariée cette framboise.}    \\ \hline
 
 \textbf{Interlocutor:} (41.69, 43.95) & \color{black} {She didn't look c- didn't look happy, it wasn't uh.}  \color{blue}  \textit{elle avait pas l'air c- pas l'air heureuse quoi c'était pas euh.} \\ \hline
 
 \textbf{Participant:} (47.59, 56.69) &  \color{black} {that's it, that's it, we would have that she had taken uh, uh that she had taken a fist in the face, *** (laugh).} \color{blue} \textit{c'est ça c'est ça on aurait que que qu'elle s'était pris euh, euh qu'elle s'était pris un poing dans la figure, *** (rire).}   \\ \hline
 \textbf{Interlocutor:} (57.01, 59.0) & \color{black} {she didn't have a black eye, did she? I don't know.} \color{blue} \textit{elle avait pas un oeil au beurre noir non ? je sais pas.}   \\ 
 \hline
\end{tabularx}
\label{tab:conversation}
\end{table}

\subsubsection{Data Availability}
The raw data employed in this work are publicly accessible. To reproduce the results of this study, only these raw data are required as an external resource. All prepossessing, preparation, and training steps are outlined in the implementation code \url{https://github.com/Hmamouche/brain_decode}. Text and conversational data can be downloaded from \url{https://hdl.handle.net/11403/convers}, and  raw fMRI signals are available in \url{https://openneuro.org/datasets/ds001740}.

\subsection{Dataset preprocessing}
\subsubsection{fMRI preprocessing}
In fMRI analysis, transforming voxels into regions of interest (ROIs) is a common step \cite{poldrack2009independence}.
This process simplifies the analysis and allows for and a better representation of raw signals and for a more explainable analysis of well investigated brain regions.
The raw fMRI datasets contain raw BOLD scans for each interaction block of each participant. Each scan is a 4D voxel image of the whole brain, with a temporal length of 385 volumes and a time-step of 1.2 seconds. This corresponds to 6 conversations, each lasting 1 minute. To extract regions of interest (ROIs) features from these scans, we use Schaefer atlas percellation \cite{schaefer2018local} using the Nilearn library \cite{Nilearn} with 200 labels for ROIs. Consequently, a masker (NiftiLabelsMasker) from the nilearn library is used to transform the raw 4D voxel scans in 200 ROIs time-series using the afordmentioned atlas parcellation.
\subsubsection{Data preparation}
To increase and enrich training data, we segmented conversations into consecutive 12 seconds splits. With a 1.2-second fMRI recording frequency, one-minute conversations yield 5 tensors, each with 10 time steps. Afterwards, for each BOLD tensor, representing a sequence of brain activity of whole brain (200 brain regions), we assign the corresponding text segment from the transcribed data of the interlocutor and the participant. This results in a dataset of 2970 samples, 250 of which were reserved for testing, and the rest were used for training (or 50 conversations for test and 544 for training).

The dataset is structured by following Visual Question Answering (VQA) datasets format using JSON files, which facilitates the integration of text, and visual modalities. Each entry consists of a dictionary with four keys:

\begin{itemize}
\item[-] \textbf{Input-text:} Text spoken by the interlocutor.
\item[-] \textbf{Bold-signal-path:} Path to the associated BOLD signal of the participant.
\item[-] \textbf{Image-path:} Path to the stimuli image.
\item[-] \textbf{Answer:}  Text spoken by the participant (the target).
\end{itemize}

Additional scripts for generating processed data from raw signals and batch loading them for training and evaluation are provided in the implementation code.

\subsection{Evaluation Metrics}
\label{subsec:metrics}
Here we present the evaluation metrics used to evaluate the performance of the models. We used three established textual measures (BLEU score, Jaccard similarity and word overlap rate) and developed a new one adapted to our task. 

\paragraph{\textbf{BLEU Score \cite{lin2004orange}:}} The BLEU score measures the quality of machine-generated translations by comparing them to reference translations. It is the product of BP (Brevity Penalty) and the geometric mean of n-gram precision. It ranges from 0 (worst) to 1 (best), with 1 indicating a perfect match with the reference.

\paragraph{\textbf{METEOR Score \cite{banerjee2004meteor}:}} The METEOR score is also a metric for  machine translation evaluation, with an improved correlation with human judgments.  It also ranges from 0 (worst) to 1 (best).

\paragraph{\textbf{Jaccard Similarity:}} The Jaccard Similarity between two sets A and B is defined as follows:
    \[ J(A, B) = \frac{|A \cap B|}{|A \cup B|} \]
In the case of text, it measures the proportion of shared words between two sentences \cite{prakoso2021short}. The value ranges from 0 (no overlap) to 1 (complete overlap).
\paragraph{Word Overlap Rate:} The Word Overlap Rate between two sets A and B is defined as a variation of the Jaccard Similarity: 
    \[ J(A|B) = \frac{|A \cap B|}{|B|} \]
It measures the proportion of elements A shares with B out of the total elements in B.
\paragraph{\textbf{Learned Perceptual Text Similarity (LPTS):}}  Given the nature of our task and the data with which we work (unstructured spoken language), we developed LPTS inspired by the Learned Perceptual Image Patch Similarity (LPIPS) (\cite{zhang2018unreasonable}), which is mainly used in image and video processing. It calculates the $l2$ distance  between the embedding vectors of two images using a pre-trained network, principally, VGGNet or AlexNet. In our case, we use a pre-trained BERT-based model for French (FlauBERT) \cite{hiebel2022clister} to generate fixed length semantically meaningful embeddings for the model's output and the ground truth then calculate the cosine difference between the two.

\subsection{Evaluated Models}

\subsubsection{Transformers benchmark}
For benchmarking, we evaluate two Transformer architectures. The first is a classic Transformer as described in \cite{vaswani2017attention} with the one difference of a linear transformation instead of the embedding layer for the encoder to account for the change in modality (text and fMRI). The second is the CNN-Transformer Hybrid Architecture as described in \cite{zhang2022cnn}.
We compare the results of three variations of the Bipartite Transformer. The first and our proposed Transformer is the Bipratite Transformer with a Deconvolution Inception module as an embedding layer alternating a Simplex, then a Duplex attention layer in the encoder and decoder blocks. The second is the same architecture with a simple linear transformation instead of the Deconvolution module. The third and final architecture is a Bipartite Transformer with a linear layer and only the duplex attention in its Transformer Blocks. These variations serve as an ablation study for our main architecture.

\subsubsection{The Multimodal LLM}
The previous transformers are compared to the proposed MLLM, which incorporates the trained encoder of the proposed Deconv-Bipratite-Transformer. In the decoding part, the architecture includes Vicuna-7B, a fine-tuned version of Llama-2  for instruction-following and conversation tasks, as described in \cite{zheng2024judging}.

\begin{table}[!t]
  \caption{Comparative results of the models evaluated on the test set.}
  \begin{tabularx}{1\textwidth}{lccccc}
    \toprule
    Model & BLEU (\%)  & METEOR (\%)  & Jaccard  & Word & LPTS  \\
     & ($\uparrow$)  & ($\uparrow$) & Similarity ($\uparrow$) & Overlap ($\uparrow$) & ($\downarrow$)  \\
    \midrule
 Classic-Transformer (Baseline) & 0.97 & 3.81 & 3.75 & 0.0394 & 0.7778  \\ \cmidrule{1-1}
 Duplex Transformer& 0.0 & 0.25 & 0.22 & 0.0023 & 0.7641 \\\cmidrule{1-1}
 CNN-Transformer& 	0.07 & 1.89 & 3.39 & 0.0347 & 0.7896  \\ \cmidrule{1-1}
 
 Bipartite Transformer & 1.54 & 3.22 & 4.42 & 0.0582 & 0.6072 \\ \cmidrule{1-1}
    
 Deconv-Bipartite-Transformer ({\textbf{Ours}})& 2.17 & 8.91 & 13.14 & 0.1668 & 0.5239  \\ \cmidrule{1-1}
        
 MLLM (\textbf{Ours})& \textbf{3.62} & \textbf{15.19} & \textbf{16.93} & \textbf{0.2835} & \textbf{0.4052}\\       
        
  \bottomrule
\end{tabularx}
  \label{tab:conv}
\end{table}

\subsection{LLM format prompting}
In VQA tasks, LLMs  tend to perform better when provided with an instruction alongside the input text \cite{li2024llava}. In our case, we use a simple, fixed prompt template as follows:
\begin{Verbatim}[commandchars=\\\{\}]
"\textcolor{red}{\{Bold tensor\}} Provide a response given this content':"
\end{Verbatim}

The MLLM receives these elements, where the textual part of the prompt are tokenized and embedded by the LLM, then concatenated with the aligned embedding of the bold signal in the same order as in the prompt. Finally, the LLM takes the concatenated embeddings as its input and generates the output text.

\subsection{Results}
During testing, we evaluate the performance of each model on each conversation within the test set. To do this, we compute the evaluations metrics, described in \ref{subsec:metrics}, that compare each predicted sentence against its corresponding ground truth sentence.
Table \ref{tab:conv} contains the average of each metric for each model. The table shows that the proposed models outperforms all other models for each metric. The Deconv-Bipartite-Transformer outperforms the other transformers, achieving BLEU and METEOR scores of (2.2\%, 8.8\%), while the second best model (Bipartite-Transformer) achieved  BLEU and METEOR scores of (1.5\%, 3.2\%). This demonstrates the effectiveness of the deconvolution and attention mechanisms employed for the proposed encoder.
The proposed MLLM outperforms significantly all evaluated models by a considerable gap in all metrics, achieving  a BLEU and METEOR scores of (3.6, 15.19\%). This result demonstrates its superior ability to translate encoded BOLD signal into text, outperforming all other models by a considerable margin in all measures.

\subsubsection{Statistical significance}
To asses the stability and the significance of the models performance, we employed a statistical testing on the BLEU scores over $10$ runs each with different weights initialization seed.  
 The Almost Stochastic Order (ASO) \cite{del2018optimal,dror2019deep} test is employed here from the deep significance Python library \cite{ulmer2022deep}. It is used to compare the performance scores of two deep learning models without making a hypothesis about the scores distribution. Given the scores of two models over multiple runs, the ASO test calculates a value $\epsilon_{min}$ that indicates how  the first model is significantly better than the second one. When $\epsilon_{min} < 0.5$ , the first model almost stochastically dominates the second . When $\epsilon_{min} = 1$ , the second model stochastically dominates the first model, and  for $\epsilon_{min} = 0.5$ , no order determined between them. 

We applied this test on the blue score of the transformer models over $10$ training/test runs with different parameters initialization seeds. The results show that the proposed Deconv-Bipartite Transformer stochastically dominates the other Transformers with $\epsilon_{min} = 1$.We did not include MLLM for this test due to computational time, but its results are stable when using different versions of Deconv-Bipartite-Transformer encoders trained using different initialization seeds.

\begin{figure}[ht]
\centering
\includegraphics[scale=0.25]{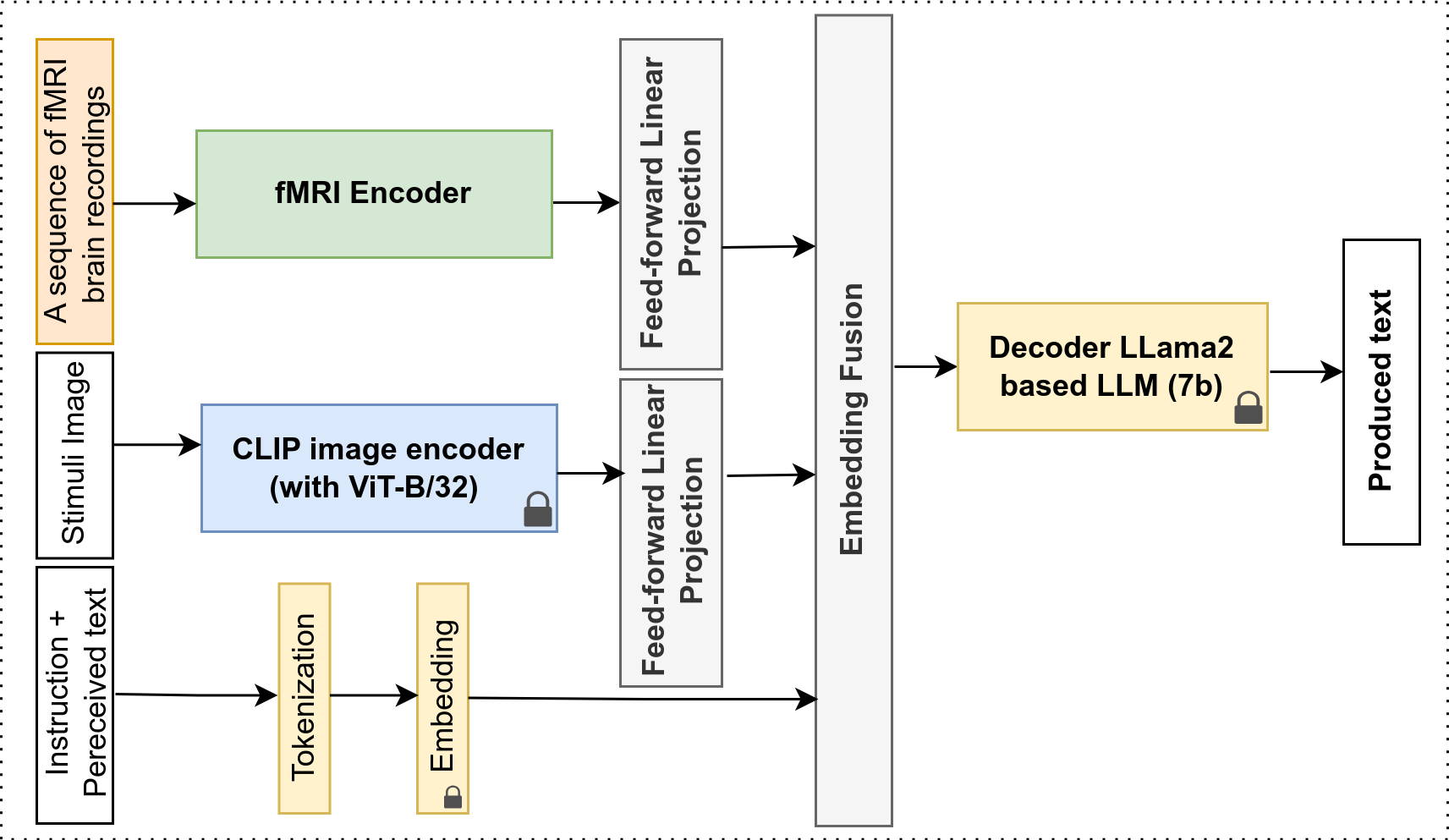}
\caption{Schema of the proposed MLLM with CLIP encoder for stimuli image integration.
.}
\label{fig:framework-v2}
\end{figure}

\subsection{Including perceived text and stimuli image}
Since the conversations between participants and interlocutors began with images of stimuli that are shown to the participant and then hidden only a few seconds before the conversations begin, it seems logical to include the perceived text and the image to the MLLM. This makes the decoding process more realistic in the sense that we get closer to how the brain perceives external information (visual and textual stimuli) and its internal state (brain activity) to generate text.

Our initial the goal of this study is to decode textual information from fMRI only. Nevertheless, we adapted our architecture to see how our model will respond to. 
For the perceived text, we added an alignment block with linear projection using a feed-forward layer similarly to the way the fixed instruction is handled in the first architecture.

For the visual information, we used a pre-trained Clip image encoder \cite{radford2021learning}, the version being based on the Vit-B/32 backbone. Finally, the embeddings of the three modalities are then fused and fed into the LLM as shown in  Figure \ref{fig:framework-v2}.

For the seek of comparison, we tested two architectures, one with two modalities, Bold signal and perceived text (MLLM-V1), and the other one with three modalities (MLLM-V2). MLLM-V0 represents the reference architecture taking as input BOLD signal only (Figure \ref{fig:framework}). 

Similarly to the previous case, the prompt template for MLLM-V1 and MLLM-V2 are respectively as follows:
\begin{Verbatim}[commandchars=\\\{\}]
"\textcolor{red}{\{Bold tensor\}} Based on this content, 
respond to the following sentence '\textcolor{red}{\{Perceived text\}}':"
\end{Verbatim}

\begin{Verbatim}[commandchars=\\\{\}]
"\textcolor{red}{\{Bold tensor, stimuli image\}} Based on this content, 
respond to the following sentence '\textcolor{red}{\{Perceived text\}}':"
\end{Verbatim}

The obtained results are presented in Table \ref{tab:conv2}. The combination of modalities yielded an performance enhancements across all metrics, notably, the inclusion of each additional modality contributed to this improvements. This suggests that the model is able to leverage perceived text visual information and generate more coherent responses.  However, the differences between MLLMs are not very significant, this can be explained by the fact that the proposed architecture is already able to capture the content of fMRI recordings, and the added modalities refine this information better.
The difference between  MLLM-V1 and MLLM-V2 is also not large. This is mainly due to the short presence of the image, as the image was been shown to the participant on time during few seconds  before the conversation began. Furthermore, the brain activity of the participant may relatively contain the visual information provided by the stimuli image, similarly, the first part of the conversation started by the interlocutor is about the image, meaning that the image should only influence the first seconds of the conversation.

\begin{table}[t]
\caption{Evaluation results of the proposed MLLM with the integration of the perceived text and the stimuli image.}
\centering
  \begin{tabularx}{1\textwidth}{lXccccc}
    \toprule

    Model & Input & BLEU (\%)  & METEOR (\%)  & Jaccard  & Word & LPTS  \\
     & Modalities & ($\uparrow$)  & ($\uparrow$) & Similarity ($\uparrow$) & Overlap ($\uparrow$) & ($\downarrow$)  \\
     
    \midrule
    MLLM-v0  & Bold signal& 3.62 & 15.19 & 16.93 & 0.2835 & 0.4052 \\
    \cmidrule{2-7}
    MLLM-v1  & Bold signal & 3.98 & 15.46 & 18.51 & 0.2966 & 0.3814\\    
      & and perceived text &  &  &  &  &  \\    
        
    \cmidrule{2-7}
     & Bold signal, &   &  &  &  &  \\  
    MLLM-v2  & perceived text & 4.1 & 17.07 & 19.39 & 0.319 & 0.3631 \\  
      &  and stimuli image &  &  &  &  &  \\      
  \bottomrule
\end{tabularx}
  \label{tab:conv2}
\end{table}

\section{Discussion}
\label{sec:discussion}
First of all, the results indicate the challenging nature of the decoding task. Given that the target text consists of spoken language from natural conversations in French, they do not contain conventional grammatical structures. Consequently, the models struggle to learn the semantic relationships between words.
Importantly, results based on the BLEU score, which is the main score used to evaluate text-to-text models, and other text similarity metrics, indicate that the proposed model is able to  generate significantly better text than the models evaluated. 
This is mainly due to the better representation the deconvolution operation gives to the fMRI data and the robustness of the bipartite attention in exploring this rich representation. The proposed MLLM performed better than all other models, validating the effectiveness of our approach. Moreover, the latest version of the proposed model performed better, which means that the proposed multimodal alignment method is able to effectively combine the three modalities and enables the LLM to generate better responses aligned with the conversation context.

Finally, this work is part of the branch of methods which aim to artificially simulate the way in which the human brain generates text from its activation.
It is important to note that decoding text from fMRI recordings will present a bootstrap for the field of neuroscience, health and rehabilitation in the coming years. For example, rehabilitation could benefit from monitoring brain activity during treatment, enabling personalized and optimized recovery plans.
This technology could also allow patients with locked-in syndrome to communicate, thereby providing access to their thoughts. In neuroscience, this will demystify language processing and memory formation, potentially leading to new non-invasive treatments.

\section{Limitations}
In this study, we demonstrated the importance of multimodal LLMs compared to classical captioning-based approaches for decoding text from brain recordings during conversations. We focused  one type of brain activity (whole brain fMRI scan), but the proposed model is flexible and can be easily generalized to other signals, such as EEG and MEG.

The main limitation of this work lies in the use of one dataset. This limitation also arises from the inherent nature of the task studied, as brain activity is not yet a common modality such as visual and textual data. This is also due to the diversity of protocols and scanners used to record brain activity, which produce signals that vary in type, format, and frequency. 
Morover, there is a scarcity of datasets that contain synchronized multimodal conversational signals and neural data during interactive tasks, given the fact that interaction context is an essential aspect when decoding text from brain activity. Unlike classic tasks such as VQA, where many datasets and models are available, researchers working on the current task often focus on specific datasets and use cases. Furthermore, collecting fMRI and other human brain data requires strict protocols and ethical considerations, which necessitates collaboration among multidisciplinary experts and scientists.

It is worth noting that the low decoding quality may have multiple causes. First, ($i$) the size of the dataset, decoding performance should normally improve with increasing data and with considering more participants. Moreover, using more standard spoken text can help the decoder generate better words since it is mainly trained using written text, while the natural spoken text of the dataset used contains words and symbols that we use orally that decoders have difficulty generating. Second, ($i$) the multi-subject aspect, which makes the task more difficult given the differences between participants' brain responses, especially since in our case we trained and tested the models on data recordings from multiple subjects, and not for each subject independently. For example, in \cite{tang2023semantic}, training and evaluation were performed per subject for a listening task, which is perhaps easier than the current task. It is important to study in a future work the decoding complexity between decoding spoken text and listening text. In this case, it might be possible to extend our experiment by adding the dataset used in \cite{tang2023semantic} and the proposed method.

Finally, decoding text from brain activity and multimodal signals is a crucial task with many future applications in healthcare and rehabilitation. However, this research field requires  more open-source datasets and comprehensive benchmarks to advance current approaches. From this perspective, it is extremely important to record new datasets with a standardized structure. For instance, for applications related to human-computer interfaces, these datasets should especially include the conversational aspect of human-human and human-robot interactions. In addition, considering multimodal signals recorded simultaneously during experiments is essential for developing more realistic and generalizable models.

\section{Conclusion}
In this study, we investigated the problem of decoding spoken text during natural conversations from fMRI recordings.
The proposed architecture is novel in terms of how it represents low-resolution and noisy fMRI signals, as well as in terms of the self-attention types incorporated for a better encoding of brain activity, and in the use of multimodal alignement learning with LLMs in an end-to-end network. To evaluate the effectiveness of our method, and given the scarcity of existing open-source implementations on this specific task, we adapted and re-implemented existing architectures from the literature and conducted a benchmark with our model.
Our proposal demonstrated superior and significant results by generating texts with better semantics compared to the ground truth. We conclude that this work will have an impact on the advancement of the state of the art of this research topic, given the potential applications that can be drawn from it, especially with the advancement of practical technologies allowing brain recording synchronized with interaction signals in real-time.

\section*{Acknowledgments}
We are grateful to the African Supercomputing Center (ASCC) of Mohammed VI Polytechnic University for providing us with the computing resources that allowed us to conduct the experiments and achieve the results presented in this article. We would like also to express our gratitude to the data collection team, particularly, Dr. Thierry Chaminade, for generously sharing the dataset \cite{rauchbauer2019brain} and making it open for research.

\bibliographystyle{unsrt}  

\end{document}